\documentclass[10pt,conference]{IEEEtran}

\usepackage{amsmath}
\usepackage{amssymb}
\usepackage{amsfonts}
\usepackage{algorithmic}

\usepackage{cite}
\usepackage[hidelinks]{hyperref}
\usepackage{cleveref}

\usepackage{graphicx}
\usepackage{textcomp}
\usepackage{xcolor}
\usepackage{booktabs}

\usepackage{colortbl}
\usepackage{soul}
\usepackage{color}

\def\BibTeX{{\rm B\kern-.05em{\sc i\kern-.025em b}\kern-.08em
    T\kern-.1667em\lower.7ex\hbox{E}\kern-.125emX}}

\definecolor{rank1}{RGB}{235,51,36}
\definecolor{rank2}{RGB}{241,90,34}
\definecolor{rank3}{RGB}{255,192,0}
\definecolor{rank4}{RGB}{146,208,80}
\definecolor{rank5}{RGB}{55,125,34}

\newcommand{\rank}[1]{
    \ifthenelse{#1=5}{\cellcolor{rank5} \color{white} 5}
    {\ifthenelse{#1=4}{\cellcolor{rank4} \color{black} 4}
    {\ifthenelse{#1=3}{\cellcolor{rank3} \color{black} 3}
    {\ifthenelse{#1=2}{\cellcolor{rank2} \color{white} 2}
    {\cellcolor{rank1} \color{white} 1}}}}
}
    
\begin{document}

\title{Software-Engineering-as-an-Artifact:\\ From Engines to Systems That Build Themselves}

\author{
    \IEEEauthorblockN{
        Julian Frattini, 
        Robert Feldt,
        Per Lenberg}
    \IEEEauthorblockA{
        \textit{Chalmers University of Technology and University of Gothenburg}\\
        Gothenburg, Sweden \\
        \{firstname\}.\{lastname\}@chalmers.se}
}


\maketitle

\begin{abstract}
    Software engineering (SE) is evolving rapidly as generative AI and LLM-based agents grow more capable.
    Yet, as a community, we risk using these new tools to solve problems that a changing SE process may no longer pose.
    We consider how AI may transform, and in part already is transforming, both the SE process and the systems it produces.
    Drawing on scenario planning methods from futures studies, we extrapolate from observed trends to a deliberately extreme scenario in which the SE process fully materializes in artifacts (like context documentation and executable skills).
    Such a process would culminate in SE engines that autonomously produce and update systems directly from requirements and user requests.
    The work of human engineers would then shift from the product to the process: they would maintain the artifacts of these engines, while the engines produce and evolve the software.
    We probe this scenario in scenario-validation interviews with four SE industry practitioners.
    The interviewees found the scenario largely plausible and desirable, and two reported that parts of industry are already moving this way; one questioned whether tacit engineering knowledge can ever be fully captured in artifacts.
    Even if this end state is reached only in part, we argue that reasoning about it can inform the intermediate stages.
    SE research should therefore take such futures into account when choosing its research directions.
\end{abstract}


\begin{figure*}[t]
    \begin{minipage}[t]{0.485\textwidth}
        \centering
        \includegraphics[width=\linewidth]{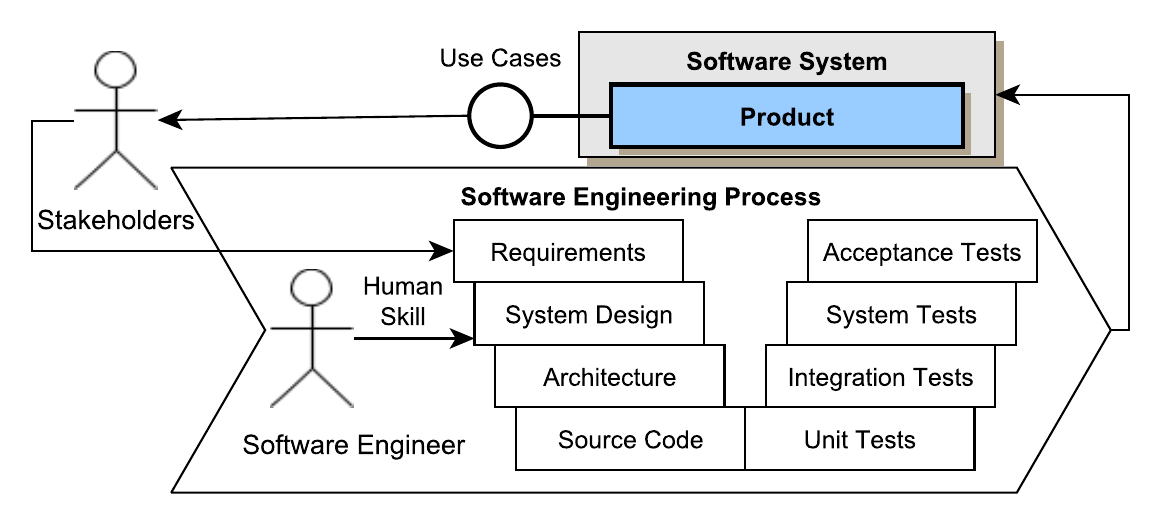}
        \caption{The Past: humans work at many abstraction levels (e.g., V-model)}
        \label{fig:se:traditional}
    \end{minipage}\hfill
    \begin{minipage}[t]{0.485\textwidth}
        \centering
        \includegraphics[width=\linewidth]{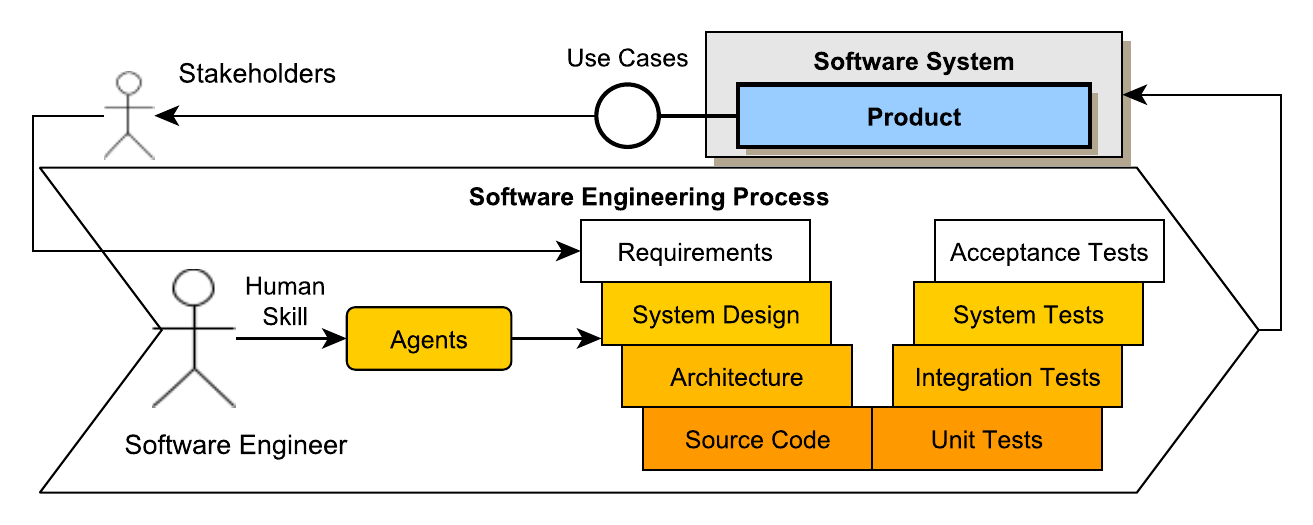}
        \caption{The Present: agents augment and automate individual tasks}
        \label{fig:se:assumed}
    \end{minipage}

    \medskip
    \begin{minipage}[t]{0.485\textwidth}
        \centering
        \includegraphics[width=\linewidth]{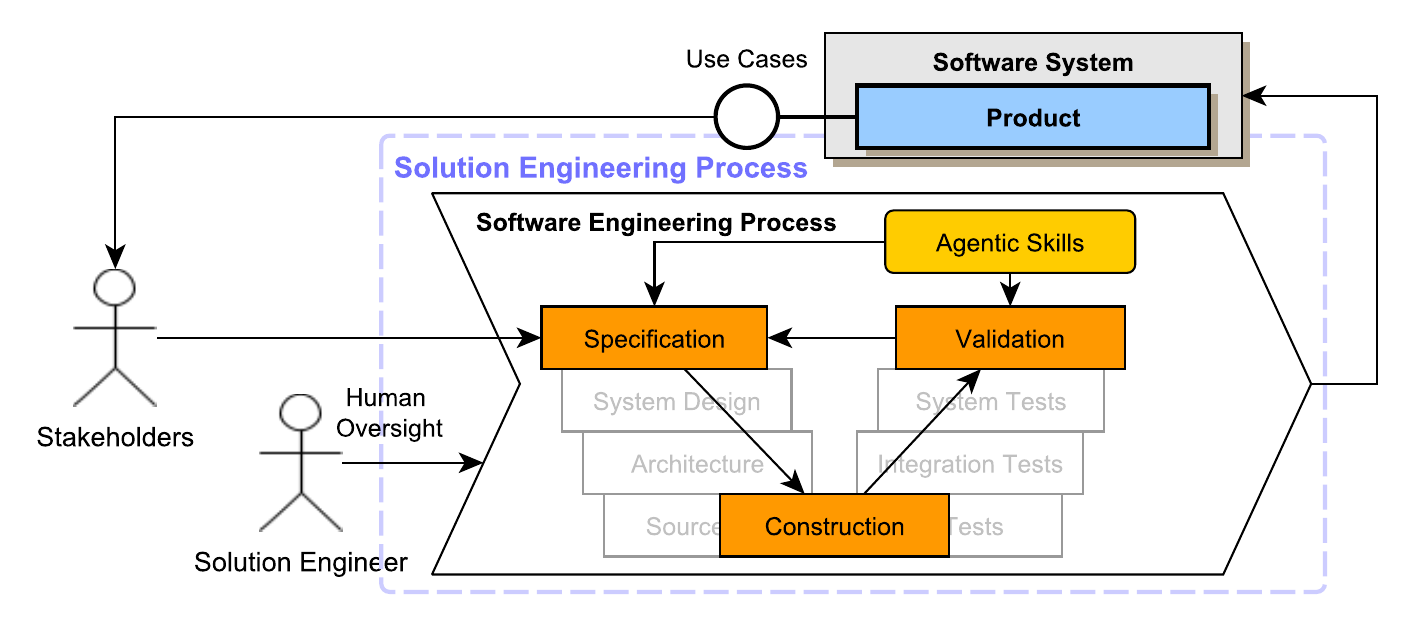}
        \caption{The Future: humans engineer the engine, not the product}
        \label{fig:se:seed}
    \end{minipage}\hfill
    \begin{minipage}[t]{0.485\textwidth}
        \centering
        \includegraphics[width=\linewidth]{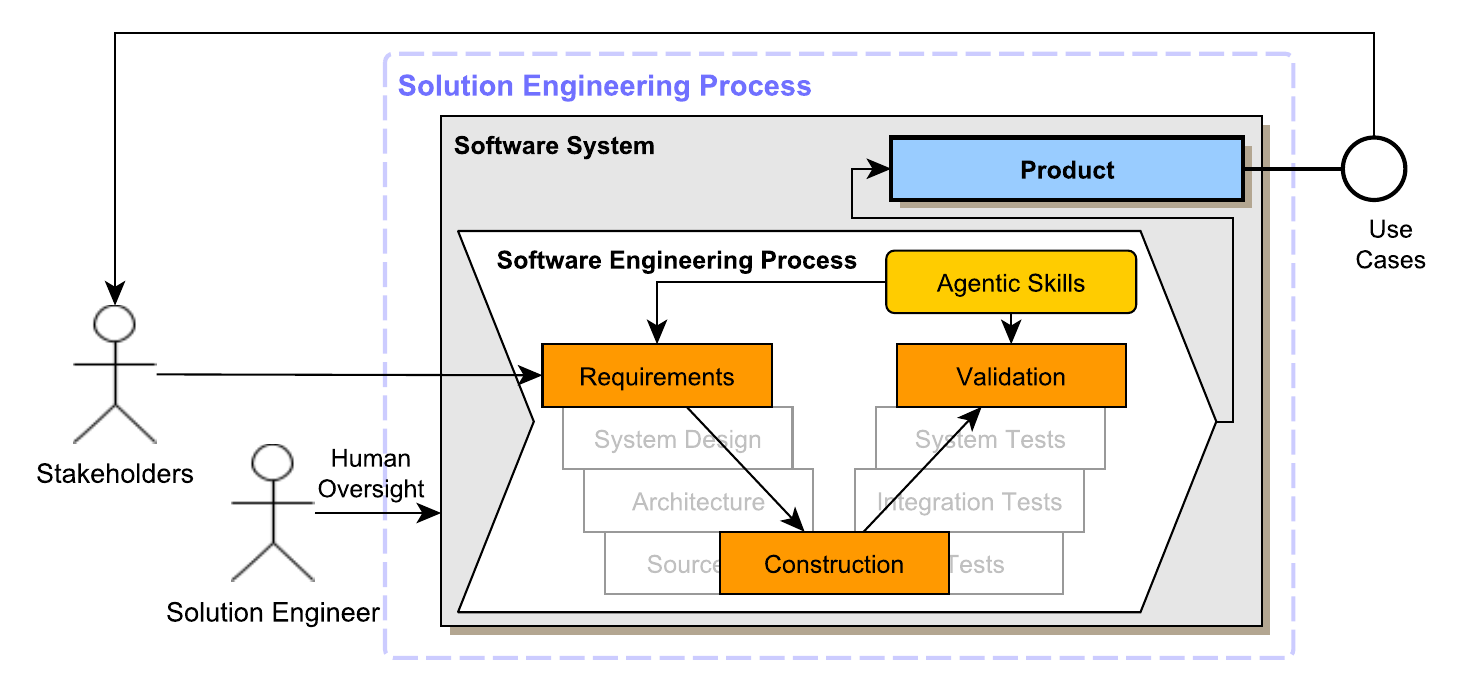}
        \caption{The End State: the engine is embedded in the product itself}
        \label{fig:se:embedded}
    \end{minipage}
\end{figure*}

\section{Introduction}
\label{sec:intro}

Software engineering (SE) is changing rapidly as artificial intelligence (AI), large language models (LLMs), and agentic systems grow more capable and widely adopted.
Researchers already observed how these technologies impact the SE process~\cite{santos2026exploring} and software products~\cite{martinez2022software}, and they anticipate further changes~\cite{terragni2025future,pezze20252030,feldt2026semi}.
Given the pace of these changes, SE may transform fundamentally, beyond isolated automation and augmentation of traditional processes.

However, many visions of the future of SE still assume that its current fundamental principles, like an iterative, incremental workflow on different levels of abstraction, remain in place.
Current and future technology may challenge this: 
The increasing automation may elevate the level of abstraction at which SE operates, i.e., contemporary artifacts like source code, unit tests, and architecture may be increasingly encapsulated in agentic workflows, leaving humans to define intent and verify output~\cite{britto2026ai}.
Ignoring this potential shift risks investing effort in research directions that may lose relevance.

SE research needs to (1) clearly articulate potential future states of SE and (2) align its research with them.
Recent visions describe how agentic SE expands the engineering object outward, from code to semi-executable artifacts, organizations, and institutions~\cite{feldt2026semi}, or how AI-native processes turn engineers into designers of environments and guardrails for agents~\cite{britto2026ai}.
We instead extrapolate the change of the SE process itself to one extreme, and we probe this scenario with practitioners.
We contribute (1) a scenario in which the SE process materializes as an engine that produces and evolves systems and ultimately moves into the product itself (\Cref{sec:vision}), (2) preliminary results from a scenario validation with four industry leaders (\Cref{sec:validation}), and (3) research directions (\Cref{sec:discussion}) and concrete plans toward this future (\Cref{sec:future}).

\section{Vision}
\label{sec:vision}

\subsection{The Past}
\label{sec:vision:past}

SE is typically understood as a disciplined and systematic approach to develop a software system that meets stakeholders' needs and constraints~\cite{ieee1983ieee}.
The goal of SE is to engineer a product which fulfills these required use cases.
\Cref{fig:se:traditional} illustrates the traditional view on SE on a high level of abstraction.
The SE process entails human software engineers applying their skills to specific tasks on different levels of abstraction.
For example, requirements engineers elicited and negotiated requirements~\cite{glinz2011glossary}, software architects partitioned the system into its parts and applied architectural patterns~\cite{perry1992foundations,clements2011documenting}, and---most prominently in SE---developers implemented functionality by creating source code.
These tasks addressed the requirements on different levels of abstraction (as indicated with the V-model in \Cref{fig:se:traditional}) to manage the complexity of the engineered product.
Note that the levels of abstraction do not imply a plan-driven or iterative software process model.

\subsection{The Present}
\label{sec:vision:present}

Generative and agentic AI have begun to shift this traditional view of SE.
The text-processing capabilities of LLMs and the autonomy of agentic systems augmented and partially even automated human tasks like coding~\cite{sergeyuk2025using} and code review~\cite{cihan2025automated}.
As indicated by the darker hues toward the bottom of the V-model in \Cref{fig:se:assumed}, this augmentation and automation starts at lower levels of abstraction and continuously escalates upwards~\cite{feldt2026semi}.
However, this perspective on SE assumes that the overall process remains largely the same:
SE continues to live in people, and these people continue to engineer the software product.
AI is assumed to change individual tasks, not the process.


\subsection{The Future}
\label{sec:vision:future}

The view in \Cref{sec:vision:present} keeps SE in the paradigm that has dominated it for decades.
We see a more fundamental shift.
Assuming that the generative, agentic technology remains economically viable and sufficiently reliable, the SE process may increasingly \emph{materialize in artifacts}~\cite{britto2026ai}: context documentation and executable skills that capture previously latent context and procedures, and can be executed by autonomous agents~\cite{feldt2026semi}.
We call this \emph{Software-Engineering-as-an-Artifact} (SE-as-an-Artifact).
In line with scenario planning, we push it to an extreme, where the process fully materializes.
Our argument does not require this extreme: it suffices that, for some types of systems and a growing share of requests, the process materializes to the degree where an engine becomes as reliable as, and cheaper than, the human process; safety-critical systems may follow later, if at all.
Knowledge that resists codification~\cite{kambhampati2021polanyi} then marks where the engine must escalate to a human, e.g., when a requirement cannot be met or behavior drifts.

The materialized SE process would then become a \emph{software engine}, i.e., a multi-agent software system that produces other software systems.
The agents within an engine operate the materialized SE process:
given stakeholders' requirements they produce new software systems.
People would then no longer engineer the software product, but the (materialized) SE process:
The same engineering principles apply, but instead of targeting the software product, they now target the process.
With similar techniques for requirements engineering, developing, and testing, human oversight would ensure that agentic skills effectively and efficiently automate the SE process.
To reflect this shift, and that an engine may deliver whichever solution meets the need, not necessarily new software, we refer to the role as \emph{solution engineers}.
That software processes can be engineered like software is not a new idea~\cite{osterweil1987software,osterweil1997software}.
What differs today is that materialized processes no longer need precise execution semantics: LLM agents execute natural-language, partially specified artifacts and fill the gaps themselves, where process programs left such gaps to humans.
This is both what enables the scenario and what limits it.

If scale economies favor a few well-maintained engines, the SE industry could contract:
instead of many dedicated SE companies, a couple of solution engineering companies would maintain their engines, and any customer could submit requirements to them to receive a system.
While this state is not yet achieved, first engines already exist for simple types of systems: prompt-to-app platforms such as Lovable or Replit Agent produce and update web applications from natural-language requests, and agents like \texttt{DroidAgent}~\cite{yoon2024intent} carry out entire SE activities such as GUI testing.

\subsection{The End State}
\label{sec:vision:endgame}

If this were achieved, it is possible that software engines, i.e., the materialized SE process, will be embedded \emph{within a software system} itself.
The resulting software system encompasses both (1) the \textbf{product}, i.e., the software system fulfilling the stakeholders' requirements, and (2) the \textbf{process}, i.e., the materialized SE process ensuring that the product meets these requirements.
\Cref{fig:se:embedded} visualizes such an embedded engine, i.e., a self-adaptive system~\cite{kephart2003vision} whose adaptation spans not only its configuration but the whole SE process, tailoring the product to stakeholders' needs immediately and thus potentially reducing time-to-market.
The remaining humans will move from \emph{in} to \emph{on} the loop: they will oversee and govern these embedded engines and handle their escalations.



\section{Validation}
\label{sec:validation}

Like any future, and all the more at today's pace of change, that of SE cannot be predicted, so a classical, positivist validation is infeasible: future scenarios cannot be falsified in the short term~\cite{amer2013review}.
To still add empirical insight, we draw on futures studies~\cite{bell2017foundations} as a ``reality check'' of our vision.

\subsection{Method}
\label{sec:validation:method}

We use \emph{scenario validation} from the scenario planning field~\cite{borjeson2006scenario,amer2013review}, which assesses properties of a scenario so that it can form an appropriate basis for decision making~\cite{chermack2001review}, as is our goal (\Cref{sec:intro}).
The following criteria are relevant~\cite{wilson1998mental}:

\begin{itemize}
    \item \textbf{Plausibility}: the scenario may possibly happen
    \item \textbf{Consistency}: the scenario is free of internal inconsistencies
    \item \textbf{Utility/relevance}: the scenario allows insights into the future that support making decisions
    \item \textbf{Challenge/novelty}: the scenario is indeed novel and challenges conventional wisdom
    \item \textbf{Differentiation}: the scenario is conceptually different from current paradigms, not a variation of the same theme
\end{itemize}

We validate the two scenarios proposed in \Cref{sec:vision:future,sec:vision:endgame} along these criteria through an interview study of leading software engineering practitioners.
Since these interviewees are potential drivers of the field toward those scenarios, we additionally validate their \textbf{desirability}~\cite{amara1981futures} as it may indicate how forcefully the scenarios will be pursued. 

\subsection{Sampling}

The target population of this scenario validation is software engineering practitioners in leading or managing roles, because we anticipate that a strategic transformation of the SE industry toward the envisioned future or end state will be driven top-down rather than bottom-up.
This target population is difficult to access, so we applied a convenience sampling method~\cite{baltes2022sampling} and invited interviewees based on availability.
We aimed for a diverse representation of sectors while ensuring that all participants had an adequate level of experience.
\Cref{tab:validation:demographics} lists the four interviewees involved in the study.

\begin{table}[hbt]
    \centering
    \caption{Demographics of Interviewees}
    \label{tab:validation:demographics}
    \begin{tabular}{l|lll}
        \toprule
        \textbf{ID} & \textbf{Sector} & \textbf{Role} & \textbf{Experience} \\
        \midrule
        A & FinTech & Head of Architecture & 20 years \\
        B & MedTech & Head of AI Transformation & 15 years \\
        C & Automotive & AI Engineering Lead & 12 years \\
        D & Telecommunication & Head of AI & 18 years \\
        \bottomrule
    \end{tabular}
\end{table}

\subsection{Process}

The first author interviewed the participants in sessions varying from 40-60 minutes.
During these sessions, the author presented the vision by describing the four stages from \Cref{sec:vision} both verbally and visually.
Then, the author invited the interviewees to assess the future and the end state on the six criteria (\Cref{sec:validation:method}), jointly unless their perception of the two differed, both on a 5-level Likert scale (e.g., from ``completely implausible'' to ``completely plausible'') and by elaborating on their assessment.
In practice, the interviewees preferred to first elaborate and then converge on a closed-ended value.

After the interview, participants filled in a questionnaire capturing their demographic data.
We elicited causal models to identify demographic factors that could bias the results, and operationalized these factors with validated instruments where possible: the current and intended AI maturity of their organization (four closed-ended questions from Alsheiabni et al.~\cite{alsheiabni2019towards}), their AI proficiency (two closed-ended questions), and their AI stance (12 closed-ended questions from Stein et al.~\cite{stein2024attitudes}).
We analyzed the interview transcripts by summarizing the closed-ended ratings descriptively and supporting each rating with verbatim statements from the interview.
When responses diverged, we triangulated them with the qualitative and the demographic data to find potential explanations.

\subsection{Results}

\Cref{tab:validation:results} shows the participants' closed-ended rating of the future scenarios on all six validation criteria.
The interviewed practitioners show an overall positive sentiment toward the vision, with several notable exceptions.

\begin{table}[hbt]
    \centering
    \caption{Closed-ended assessment of the six validation criteria (split by Future/End State where they diverged)}
    \label{tab:validation:results}
    \begin{tabular}{l|ccccc}
        \toprule
        \textbf{Participant}: & \textbf{A} & \textbf{B} & \textbf{C} & \multicolumn{2}{c}{\textbf{D}} \\
        \textbf{F}uture/\textbf{E}nd State: & & & & F & E \\
        \midrule
        Plausibility & \rank{5} & \rank{5} & \rank{2} & \multicolumn{2}{c}{\rank{5}} \\
        Consistency & \rank{5} & \rank{4} & \rank{5} & \multicolumn{2}{c}{\rank{5}} \\
        Utility/Relevance & \rank{5} & \rank{3} & \rank{2} & \multicolumn{2}{c}{\rank{5}} \\
        Challenge/novelty & \rank{3} & \rank{5} & \rank{4} & \rank{2} & \rank{5} \\
        Differentiation & \rank{5} & \rank{5} & \rank{4} & \multicolumn{2}{c}{\rank{5}} \\
        Desirability & \rank{3} & \rank{5} & \rank{5} & \multicolumn{2}{c}{\rank{5}} \\
        \bottomrule
    \end{tabular}
\end{table}

Among the less controversial criteria, the internal consistency and the differentiation of the visions from the past and present were uniformly supported by the interviewees.
The perception of the other criteria diverged.
Participant C was particularly skeptical of the plausibility of the vision. 
While they agree that ``software engineering will become automatic in some way''~(C) they see a substantial barrier to the manifestation of latent skills in artifacts, as implied in \Cref{sec:vision:future}.
In their view, the natural language interface with which solution engineers can instruct agents to automate the SE process is far less trivial than currently anticipated.
In reference to Michael Polanyi's dictum ``we know more than we can tell''~\cite{kambhampati2021polanyi}, they are convinced that there are certain latent concepts that are impossible to materialize.
One symptom of this is the emergence of \emph{taste} in AI-driven SE or ``vibe-coding'', capturing latent and non-materializable preferences of end users in a software system~\cite{liu2026syntax}.
According to participant C, purely language-based AI may not suffice for the vision.

On the other hand, participants A and D reported that this vision was barely novel in some parts of SE industry.
According to A, for ``the new items that we are building, [...] it's a convention''~(A) to build them this way.
For them, ``the ultimate goal is to actually move people from be[ing] \emph{in} the loop to be[ing] \emph{on} the loop''~(D) where the engine ``should be as reliable as a compiler is''~(D).

Participants B and C were also more critical of the utility and relevance of the vision.
Participant B settled at 3: the vision may repel many practitioners and researchers committed to traditional SE values, while industry overall may find it a very useful guideline.
Participant C sees the ``pathway as being most relevant instead of [...] the end state of the future software engineering process''~(C).

Finally, participant A rated desirability 3: the vision is ``not desirable in the sense of human being and loss of jobs''~(A), yet ``it's a completely desirable solution, perhaps for humanity''~(A).
This tension reflects two forces acting on SE: engineers anxious about their expertise and jobs versus the market's drive for better-fitting software, delivered faster.
The participants share this view, e.g., ``99\% of all people don't care if it's software or whatever that solves their problem''~(B).

\section{Discussion}
\label{sec:discussion}

\subsection{Implications}

The scenario validation indicates that practitioners take this future seriously and that some forces already move toward it, while codifiability remains contested.
Based on the interviews, we see four research directions.
\emph{Codifiability of skills}: Some latent skills or context may resist materialization in artifacts and exceed what models have learned, which would limit what an SE engine can automate~\cite{kambhampati2021polanyi}; we need methods to determine what can and should be codified, to elicit and materialize it, and ways of working in which engines escalate the rest to humans.
\emph{Certification of generated software}: When source code becomes a lower-level artifact, produced in masses yet hardly understood by solution engineers (like byte-code and assembly today), new techniques must certify it against requirements and regulations~\cite{britto2026towards}.
\emph{Human--engine interfaces}: Where natural language cannot comprehensively express requirements, engines need other modalities~\cite{kambhampati2021polanyi}; and as end users request changes directly from (embedded) engines, these requests must be clarified, reconciled across users, and governed.
\emph{Workforce integration}: If SE contracts to a small number of highly specialized solution engineers, this will, as in previous industrial revolutions, rightfully evoke resistance; this transition must be transparent, fair, and aware of human factors~\cite{tenakwah2025embracing}.
Beyond the interviews, the vision raises a fifth direction, \emph{engineering the engine}: if defects are fixed in the process rather than the product, skills and context artifacts must be tested, debugged, and versioned as rigorously as code, and product failures traced to the skill that caused them.

\subsection{Limitations}

The scenario validation is a reality check, not a test of the vision, and has several limitations.
Our target population was limited to managers in the SE context, who will view the vision differently from other roles, particularly those whose profession may be automated.
Since these managers drive change in SE practice, we deem the target population appropriate, but three of the four hold AI-specific roles and may lean toward AI-driven change.
The first author both presented the scenarios and elicited the ratings in the same session, which may inflate ratings; the participants' reasoning, and especially their dissent, is therefore more informative than the scores.
Participants assessed the scenario before the qualifications in \Cref{sec:vision:future} were added.
Likewise, plausibility (``may possibly happen'') is a deliberately low bar: the ratings show that the scenario was not dismissed, not that it is likely.
Further, we did not aim for generalizability but for rich, exploratory insights; the four participants' generally positive view comes with varied concerns.
Finally, the insights are self-reported perceptions, since the envisioned future cannot yet be observed.
Our replication package~\cite{replicationpackage} contains the interview slides and each participant's ratings with their verbatim justifications; raw data is shared only where participants consented.

\section{Future Plans}
\label{sec:future}


This future is neither certain nor near.
Yet our interviewees found it largely plausible and partly desirable, so the technical and human challenges on the way there deserve study.
All four participants, even C, want to explore the next steps with us.
We will first prototype the SE engine of \Cref{sec:vision:future} for a bounded class of systems.
The prototype lets us probe C's objection: which latent skills can be materialized in natural-language artifacts, and how reliably does the engine perform each SE activity?
We also plan to study how the workforce responds to increasing automation, to describe the solution engineer's skill set, i.e., the human skills that remain relevant, and guide retraining toward it.
We will track signposts of the scenario, e.g., whether organizations test and version skills like code and fix defects in skills rather than products.
As participant B put it: ``If we believe that this future is plausible, we should give the people involved a heads-up.''~(B)

\section{Conclusion}
\label{sec:conclusion}

Stakeholders, including end users, do not necessarily want software; they want their needs and requirements met.
As a discipline, we must accept that software is only one possible solution.
If SE research keeps studying yesterday's process, it risks becoming irrelevant to practice.
We should instead track where our field may be heading and choose research directions accordingly.
SE-as-an-Artifact is one such possible future: SE becomes part of a broader \emph{solution engineering}, its principles apply to the process, materialized as an engine that builds a solution, rather than to the product.
Ultimately, that engine may move into the product itself.

\bibliographystyle{IEEEtran}
\bibliography{references}

\end{document}